\documentclass[iop]{emulateapj}

\usepackage{amsmath}
\usepackage{natbib}
\shorttitle{HS1700}
\shortauthors{Syphers et al.}

\begin{document}

\title{The \ion{He}{2} Post--Reionization Epoch: {\it HST}/COS Observations of the Quasar HS1700+6416}

\author{David Syphers and J.\ Michael Shull }
\affil{CASA, Department of Astrophysical and Planetary Sciences, University of Colorado, Boulder, CO 80309, USA}

\email{david.syphers@colorado.edu, michael.shull@colorado.edu}  

\begin{abstract}

The reionization epoch of singly ionized helium (\ion{He}{2}) is believed to start at redshifts $z \sim 3.5$--4 and be nearly complete by $z \simeq 2.7$.
We explore the post-reionization epoch with far-ultraviolet spectra of the bright, high-redshift quasar HS1700$+$6416 taken by the Cosmic Origins Spectrograph (COS) on the {\it Hubble Space Telescope}, which show strong \ion{He}{2} ($\lambda 303.78$) absorption shortward of the QSO redshift, $z_{QSO} = 2.75$.
We discuss these data as they probe the post-reionization history of \ion{He}{2} and the local ionization environment around the quasar and transverse to the line of sight.
We compare previous spectra taken by the {\it Far Ultraviolet Spectroscopic Explorer} to the current COS data, which have a substantially higher signal-to-noise ratio.
The Gunn--Peterson trough recovers at lower redshifts, with the effective optical depth falling from $\tau_{\rm eff} \simeq 1.8$ at $z \sim 2.7$ to $\tau_{\rm eff} \simeq 0.7$ at $z \sim 2.3$.
We see an interesting excess of flux near the \ion{He}{2}~Ly$\alpha$ break, which could be quasar line emission, although likely not \ion{He}{2}~Ly$\alpha$.
We present spectra of four possible transverse-proximity quasars, although the UV hardness data are not of sufficient quality to say if their effects are seen along the HS1700 sightline.

\end{abstract}


\section{INTRODUCTION}
\label{sec:intro}

The full reionization of intergalactic helium, from \ion{He}{2} to \ion{He}{3}, was likely delayed compared to that of hydrogen because of the need for higher-energy photons than stars provide.
While hydrogen was likely reionized at $z>6$ \citep[e.g.,][]{fan06,shull12}, quasars may have created sufficient photons with $E>4$~ryd to reionize helium at $z \sim 3$--4 \citep{furlanetto08}.
This is consistent with redshift estimates from the intergalactic medium (IGM) temperature \citep[e.g.,][]{becker11}, which increases noticeably during helium reionization, as well as estimates from the redshift evolution of the \ion{He}{2} Gunn--Peterson optical depth (e.g., \citealt{syphers11a,syphers12,worseck11a}; but see \citealt{davies12}).

The Gunn--Peterson test, in which a non-trivial ion fraction creates a trough by line absorption at every redshift \citep{gunn65}, is the most direct test of the later stages of helium reionization.
Unlike for hydrogen, this test is useful for helium even well before the completion of reionization, up to a \ion{He}{2} fraction $x_{\rm HeII} \sim 0.1$ \citep{mcquinn09a,syphers11b}, and indicates that \ion{He}{2} reionization completes at $z \sim 3$ or just below.
There is noticeable sightline variance, however, and some sightlines have $x_{\rm He II} \gg 0.01$ down to redshift $z=2.7$ \citep{shull10}.

HS1700+6416 (henceforth HS1700) is an extremely luminous ($r=15.9$) quasar at redshift $z_{\rm QSO} = 2.75$.
Despite having several partial Lyman-limit systems \citep[pLLS;][]{vogel95}, it is still bright in the far UV, with flux down to the \ion{He}{2} Ly$\alpha$ break at 303.78~\AA\ in the rest frame \citep{davidsen96}.
This makes it a ``\ion{He}{2} quasar,'' one of the few percent of $z \sim 3$ quasars that have visible flux at the \ion{He}{2} Ly$\alpha$ break owing to unusually low integrated \ion{H}{1} column densities along their sightlines.

In recent years, cross-matching quasar catalogs with {\it Galaxy Evolution Explorer} \citep[{\it GALEX};][]{morrissey07} UV imaging has lead to a vast increase in the number of known \ion{He}{2} quasars \citep{syphers09b,syphers09a,syphers12,worseck11a}.
However, \ion{He}{2} quasars that are ``very bright,'' with far-UV flux $f_{\lambda}^{\rm FUV} \gtrsim 10^{-15}$~erg~s$^{-1}$~cm$^{-2}$~\AA$^{-1}$, remain very rare.
The only three known, in order of brightness, are HE2347$-$4342 \citep{reimers97,shull10}, HS1700, and 4C57.27 \citep{syphers12}.
HS1700 is the lowest redshift of these, and with no black Gunn--Peterson trough, it probes the helium post-reionization epoch.

In Section~\ref{sec:obs} we present a new {\it Hubble Space Telescope} {\it (HST)} far-UV spectrum of HS1700, and consider instrumental effects that might impact the science.
In Section~\ref{sec:tau} we calculate the Gunn--Peterson optical depth seen in the HS1700 sightline, the subject of many previous observations, and we discuss the most relevant of these in Section~\ref{sec:comparison}.
Although the new observations presented are superior in some respects, there are resolution and wavelength coverage advantages in some of the older data sets, which allow us to better interpret the current data.
We discuss possible quasar line emission near the \ion{He}{2} break in the context of comparison to the {\it Far Ultraviolet Spectroscopic Explorer} {\it (FUSE)} spectrum.
\ion{He}{2} observations, when combined with \ion{H}{1} spectroscopy, uniquely allow determination of the UV background hardness throughout most of the IGM, and we discuss this in Section~\ref{sec:eta}.
In Section~\ref{sec:proximity} we discuss proximity effects, both from HS1700 itself and also transverse proximity effects from lower-redshift ionizing sources.
We conclude in Section~\ref{sec:conclusion}, and include a detailed discussion of our modeling of the spectrum background in the Appendix.

\section{OBSERVATIONS}
\label{sec:obs}

\begin{figure}
\epsscale{1.2}
\plotone{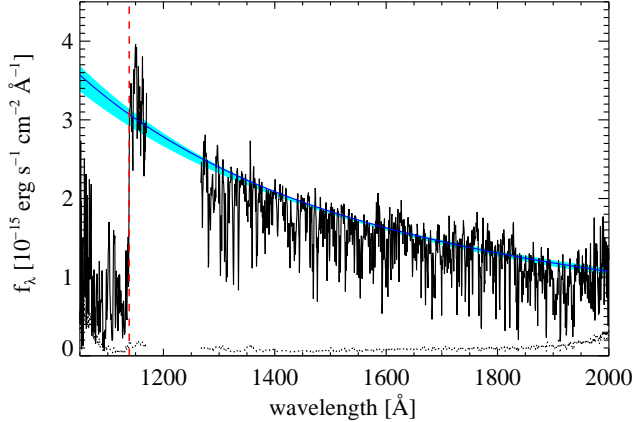}
\caption{The COS G140L spectrum of HS1700, with a power-law continuum fit overplotted in blue, and continuum uncertainty in cyan. The COS data are binned to $0.56$~\AA\ (about one resolution element). The dashed vertical red line shows the expected wavelength of the \ion{He}{2}~Ly$\alpha$ break based on the quasar redshift, $z=2.748 \pm 0.005$. Continuum fits including partial Lyman-limit systems are discussed in Section~\ref{sec:stis_comp}, and shown in Figure~\ref{fig:extrapolated_continua}.}
\label{fig:full_cos}
\end{figure}

HS1700 was observed for 15,705~s on October 10, 2011, using the Cosmic Origins Spectrograph \citep[COS;][]{green12} aboard {\it HST}.
We used the G140L grating (resolving power $R \simeq 2000$--3000) to observe to wavelengths as short as possible.
The spectrum is shown in Figure~\ref{fig:full_cos}, including a continuum fit discussed in Section~\ref{sec:tau}.
The data were taken in TIME-TAG mode, which allowed us to make a cut avoiding those times with evident geocoronal \ion{O}{1}~$\lambda$1304, resulting in 5900~s of data during orbital night.
In our plots and analysis, night-only data are used in wavelength regions contaminated by geocoronal emission, while all data are used elsewhere, because there is no evidence for a noticeable scattered-light background from geocoronal lines \citep{syphers12}.
The data are processed with CALCOS 2.18.5, using a restricted extraction window, pulse height amplitude cuts, and an improved background subtraction method.
These modifications of the standard extraction method are described in detail in the Appendix; they noticeably affect only data where the background contributes significantly to the spectrum ($\lambda \lesssim 1100$~\AA).

We also require optical data to examine the \ion{H}{1} corresponding to \ion{He}{2} absorption, and for this we use the Keck HIRES spectrum from \citet{simcoe02}.
This observation gave data with a resolution of 6.6~km~s$^{-1}$ and a median S/N of 110 per resolution element.

The redshifts quoted in the literature for HS1700 vary significantly.
At $z \sim 3$, the prominent quasar lines seen in optical spectra are high-ionization UV emission (including \ion{H}{1} Ly$\alpha$ and \ion{C}{4}~$\lambda$1549), which can be shifted by many hundreds or even thousands of km~s$^{-1}$ from the systemic redshift of the quasar \citep[e.g.,][]{vanden-berk01}.
Because low-ionization metal lines tend to trace the systemic redshift much better, we measure \ion{O}{1}~$\lambda$1304 and \ion{C}{2}~$\lambda$1335 to find $z_{\rm QSO}=2.745 \pm 0.004$ (using the [\ion{O}{3}]--corrected rest wavelengths of \citealt{vanden-berk01}).
\citet{trainor12} adopted the slightly higher redshift $z_{\rm QSO}=2.751 \pm 0.003$ based on \ion{Mg}{2}~$\lambda$2798 in a Palomar/TripleSpec spectrum and H$\gamma$ in a Keck/NIRSPEC spectrum.
(The best lines for quasar redshift determination, H$\beta$ and [\ion{O}{3}]~$\lambda$5007, are unfortunately blocked by atmospheric absorption at this redshift.)
The \ion{He}{2} break is consistent with both of these redshifts, particularly given the wavelength calibration uncertainty we discuss in section~\ref{sec:wavcal}.
The difference between the two redshifts is only $1.2$$\sigma$, so we adopt the average $z_{\rm QSO}=2.748 \pm 0.005$ for this paper.
The error is unfortunately not easily quantified, given the systematic uncertainties involved.
Our adopted error is slightly larger than the statistical error in accommodation of this.
We note that small redshift variations have no impact on our analysis of the \ion{He}{2} Ly$\alpha$ forest, and matter only for discussion of the line-of-sight proximity effect (Section~\ref{sec:LOS_prox}).
The onset of the hydrogen Ly$\alpha$ forest at $z=2.7442$ is consistent with our adopted redshift, and clearly rules out the significantly lower $z_{\rm QSO} \simeq 2.72$--$2.73$ used in some earlier works \citep[e.g.,][]{reimers93,fechner06,wu10}.

\subsection{COS Wavelength Calibration}
\label{sec:wavcal}

The G140L/1280 central wavelength setting uses both segments of the COS FUV detector, with a substantial gap in between.
Segment~A covers $\lambda \gtrsim 1270$~\AA, while segment~B covers $\lambda \lesssim 1170$~\AA.
COS wavelength calibration is accomplished using the spectrum of an onboard PtNe lamp, but there are no useable lines produced on segment B for G140L.
Therefore the segment B wavelength solution is taken entirely from segment A \citep{oliveira10}.
We have checked the wavelength calibration of COS segment~A using strong ISM absorption lines, which show offsets of $\sim$30~km~s$^{-1}$, substantially smaller than a resolution element (the latter is $\simeq$100--150~km~s$^{-1}$).

Segment B does not appear so well aligned.
The only strong ISM lines on segment~B outside the \ion{He}{2}~Ly$\alpha$ forest are from \ion{Fe}{2}, the strongest of which should be \ion{Fe}{2}~$\lambda$1144.94~\AA.
This line appears to be observed at $1142.96 \pm 0.05$~\AA; although there are other \ion{Fe}{2} lines closer to this wavelength, no absorption lines are evident in the spectrum at longer wavelength.
Another method of determining a wavelength offset is to cross-correlate the \ion{He}{2} and \ion{H}{1} forests, because IGM density fluctuations should on average connect the two.
A clear peak in the cross-correlation also suggests that the UV data need to be shifted to higher wavelengths, but by about 3.4~\AA\ rather than 2.0~\AA.
Unfortunately, the low resolution and modest S/N of the COS data preclude a firm determination of the wavelength offset between optical and UV data.

Cross-correlation of COS data with the independently calibrated {\it FUSE} spectrum also gives a strong peak at a 2.0~\AA\ shift of the COS data, and the {\it FUSE} spectrum shows ISM \ion{Fe}{2}~$\lambda$1143.2~\AA\ and 1144.9~\AA\ to be within $0.1$--$0.2$~\AA\ of their expected locations.
This magnitude of shift also makes the \ion{He}{2} break most consistent with the adopted redshift.
We therefore use the COS segment~B spectrum shifted by $2.0$~\AA\ as the standard version in our plots and discussion, but caution that this is not an entirely firm determination. 

\section{EVOLUTION OF \ion{He}{2} EFFECTIVE OPTICAL DEPTH}
\label{sec:tau}

In order to determine the optical depth of IGM \ion{He}{2}, we need to extrapolate the quasar continuum to lower wavelengths.
\citet{syphers12} found that variations in smoothing length and other parameters dominate the error in this extrapolation of COS data.
We follow their method of varying these values and taking the resulting range of continua as our uncertainty.
We fit a power law to the COS FUV data, adopting $E(B-V)=0.026$ \citep{schlegel98}, $R_V=3.1$, and the extinction curve of \citet{fitzpatrick99}.
The COS spectrum, with its estimated continuum, is shown in Figure~\ref{fig:full_cos}.
Small uncertainties in the reddening are negligible, since, for example, using $E(B-V)=0.022$ \citep{schlafly11} changes our observed continuum estimation by only 2\% at 1000~\AA.
Larger uncertainties in the reddening, $R_V$, or the extinction curve are systematics we do not account for in our errors.

While all we require for calculating $\tau_{\rm eff}$ is an extrapolation of the continuum below the \ion{He}{2} break, HS1700 is one of the very few quasars for which we have continuous spectral coverage from the FUV to the optical.
It is thus possible to actually measure the intrinsic power-law index, taking into account the many pLLS.
This does not greatly affect the extrapolated values, but is of interest for determining the true extreme-UV (EUV, $\lambda \lesssim 912$~\AA) spectral index.
The small impact on $\tau_{\rm eff}$ discussed in detail in Section~\ref{sec:stis_comp}.

Detector background is important whenever the background counts are significant compared to the source counts.
This happens with faint targets, high optical depth regions, and---most relevant to HS1700---at wavelengths where the instrumental throughput is very low.
We are little affected by the background uncertainty at $\lambda \gtrsim 1100$~\AA\ (\ion{He}{2} Ly$\alpha$ redshift $z \gtrsim 2.62$), but at shorter wavelengths {\it HST}/COS has very poor throughput, dropping from an effective area of nearly 2000~cm$^2$ at 1170~\AA\ to only 100~cm$^2$ at 1090~\AA\ and 10--20~cm$^2$ at $\lambda < 1070$~\AA\ \citep{mccandliss10}.
The default CALCOS background subtraction is problematic for a number of reasons, notably the problem of ``$y$-dip,'' where reduced sensitivity in the primary science aperture (PSA) on the detector leads to a lower background in the PSA compared to the region where the background is measured.
For a substantially more detailed discussion of the COS background and the way CALCOS handles it, see \citet{syphers12}.
We have dramatically improved background determination for COS data, and discuss our method in detail in the Appendix.

\begin{figure}
\epsscale{1.2}
\plotone{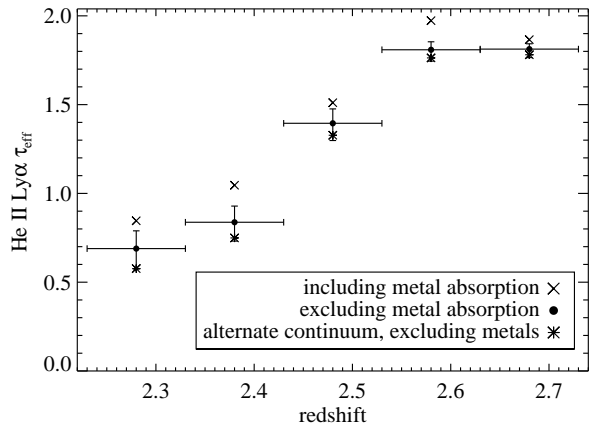}
\caption{The broad-scale evolution of the \ion{He}{2}~Ly$\alpha$ effective optical depth with redshift along the HS1700 sightline. We show confidence intervals on $\tau_{\rm eff}$ including both 68\% Poisson uncertainty on observed counts and the range of possible extrapolated continua shown in Fig.~\ref{fig:full_cos}. We account for metal contamination using the model of \citet{fechner06a}, as discussed in the text. The alternate continuum includes many pLLS, as discussed in Section~\ref{sec:stis_comp} and shown in Figure~\ref{fig:extrapolated_continua}.}
\label{fig:tau_evolution}
\end{figure}

One additional concern for the calculation of $\tau_{\rm eff}$ is that by necessity one usually assumes that {\it all} absorption is due to \ion{He}{2}.
Hydrogen contamination is unlikely to be a major concern, as only weak higher-order lines of relatively rare low-redshift systems would affect this spectral region, but metal absorption can be noticeable.
HS1700 is again fairly unusual among \ion{He}{2} quasars because it has a very well-studied high-resolution optical spectrum, and much effort has gone into modeling its metal systems.
\citet{fechner06a} performed detailed modeling of the HS1700 sightline, predicting the metal-line spectrum seen in the far-UV based on observed absorption systems and Cloudy modeling.
Accounting for metal contamination is important when doing line fitting of the FUV spectrum \citep{fechner06}, but it does not have a large effect when finding the effective optical depth over large bins.
We nonetheless do account for their predicted metal spectrum \citep[Fig.~7 in][which also includes Galactic H$_2$]{fechner06a}, as it can shift $\tau_{\rm eff}$ by $0.1$--$0.15$ overall, and more in some redshift bins.
(Their model does not cover $z < 2.292$ or $2.562 < z < 2.578$ since it was made to cover {\it FUSE} LiF data only.)
However, we caution that such predictions are not possible for many \ion{He}{2} quasars; we therefore show $\tau_{\rm eff}(z)$ in Fig.~\ref{fig:tau_evolution} with and without modeling this absorption.
Theoretical models do not yet depend on such small shifts, however.

The large-scale evolution of the \ion{He}{2} Ly$\alpha$ optical depth is shown in Figure~\ref{fig:tau_evolution}.
As one would expect, the average optical depth decreases toward lower redshift.
The COS data show a noticeably higher $\tau_{\rm eff}$ than was calculated for the {\it FUSE} data in \citet{fechner06}; the COS data on HS1700 give average optical depths much more line with those seen in HE2347 at similar redshifts \citep{shull10}.
Note that many previous \ion{He}{2} studies picked specific and irregular bins to capture what they felt were structures in $\tau_{\rm eff}(z)$ \citep[e.g.,][]{heap00,zheng04b,shull10}.
Although theoretical models have been compared to these values for want of any other data \citep[e.g.,][]{dixon09}, we feel this practice is not useful in general, and bins should be regular and made without regard to specific sightline structure to enable comparison to models (which, after all, are trying to reproduce average behavior, not specific structures).
However, this specifically addresses broad $\tau_{\rm eff}(z)$ evolution measurements.
A promising method that {\it should} examine regions chosen {\it a posteriori} is using ``dark gaps'' for helium the way it has been done for hydrogen \citep[e.g.,][]{paschos05,gallerani06}.

Our uncertainties incorporate the full range of acceptable power-law fits to the data, but these fits do not include any pLLS.
If we consider the best-fit continuum including pLLS, the $\tau_{\rm eff}$ values are slightly lower, as shown by the asterisks in Figure~\ref{fig:tau_evolution}.
Such a continuum fit would marginally strengthen the case for increasing $\tau_{\rm eff}$ at higher redshift, but in every case the points lie within the 68\% confidence intervals for the points from pure power-law fits.

In addition to broad evolution of the optical depth, the detailed small-scale evolution of $\tau_{\rm eff}(z)$ tells about the IGM density and local ionizing sources.
We consider this in Section~\ref{sec:eta}.

\section{PRIOR OBSERVATIONS AND POTENTIAL LINE EMISSION}
\label{sec:comparison}

As one of the FUV-brightest high-redshift quasars on the sky, HS1700 has been observed many times before.
It is bright enough to have been observed with the {\it International Ultraviolet Explorer} \citep[IUE;][]{reimers89}.
It was observed early with {\it HST} using the Faint Object Spectrograph \citep[FOS;][]{reimers93}, which did not have coverage down to the \ion{He}{2}~Ly$\alpha$ break, but showed that it had flux extending though the FUV.
A {\it Hopkins Ultraviolet Telescope} {\it (HUT)} observation confirmed that it was a \ion{He}{2} quasar \citep{davidsen96}.
Subsequent observations with the Goddard High-Resolution Spectrograph (GHRS) and the Space Telescope Imaging Spectrograph (STIS) were not put to use for helium-related science, but the {\it FUSE} spectrum (calibrated by low-resolution STIS) made it one of only two quasars to date with resolved \ion{He}{2}~Ly$\alpha$ forest observations.
STIS covered the FUV and NUV, thereby allowing a fit of the entire spectrum, and the {\it FUSE} observations remain the highest resolution to date, albeit at low S/N.

To begin we make a brief comparison to an older data set, the FOS spectrum of \citet{vogel95}.
We recommend that all FUV line identifications in this paper be confirmed in the COS spectrum, which has twice the resolution and higher S/N.
The FOS spectrum may be suspect in some areas, with systematics larger than the quoted uncertainties.
For example, the \ion{He}{1}~$\lambda$584 absorption line reported at $z=2.1678$ \citep{reimers93,vogel95} does not exist in the COS spectrum, although the neighboring Galactic \ion{Al}{3}~$\lambda$1854 clearly does.
Locations for claimed \ion{He}{1}~$\lambda$584 at $z=2.290$ and $z=2.315$ do show real lines in the COS spectrum, which is important, as HS1700 is the only sightline with confirmed detections of this line in the IGM.

\subsection{Possibility of EUV Quasar Emission Lines}
\label{sec:line_emission}

\begin{figure}
\epsscale{1.2}
\plotone{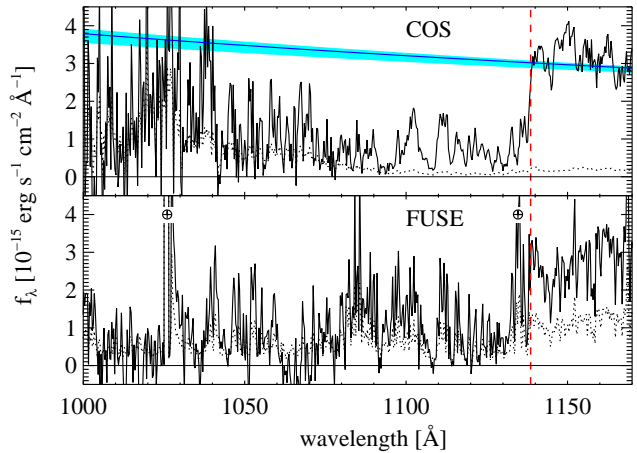}
\caption{The \ion{He}{2}~Ly$\alpha$ forest along the HS1700 sightline as seen in COS and {\it FUSE}. The COS data are binned to $0.32$~\AA\ (just over half a resolution element), and the {\it FUSE} data are binned to $0.30$~\AA\ (about six resolution elements). The {\it FUSE} data are {\it not} convolved with the COS line-spread function \citep[LSF;][]{kriss11}; one can therefore see the advantage of higher resolution, in particular being able to identify small black regions. The COS data show the substantial S/N improvement this detector has at longer wavelengths, as well as the much better characterized background. The dashed vertical red line shows the expected wavelength of the \ion{He}{2}~Ly$\alpha$ break based on the quasar redshift. Geocoronal Ly$\beta$ and \ion{N}{1}~$\lambda$1134 contamination is marked in the {\it FUSE} spectrum.}
\label{fig:cos_fuse}
\end{figure}

One puzzle in EUV quasar spectra has been a lack of clear \ion{He}{2}~Ly$\alpha$ emission in most quasars, despite clear photoionization model predictions that this line should be evident in the broad line region \citep{syphers11a,syphers12,lawrence12}.  
The first claim of clearly seeing this line was \citet{davidsen96} using {\it HUT} data on HS1700.
However, no such line was seen in {\it FUSE} data (Figure~\ref{fig:cos_fuse}).
It seems clear from Figures~\ref{fig:full_cos} and \ref{fig:cos_fuse} that no reasonable continuum fit easily explains the excess of flux near the \ion{He}{2}~Ly$\alpha$ break.
It is unfortunate that the COS spectrum has a gap just redward of this rise in flux, allowing for the possibility that flux misalignment is the source of the problem (albeit a very unlikely one, since flux matching across segments has not been problematic with other data).

A flux excess is seen above the continuum fit at $1150 \pm 2$~\AA, approximately 11~\AA\ redward of the expected location of \ion{He}{2} Ly$\alpha$.
In the rest frame, the peak occurs at $306.8 \pm 0.5$~\AA\ ($\sim$3000~km~s$^{-1}$ redward of \ion{He}{2} $\lambda 303.78$).
Quasar broad emission lines are often notably blueshifted from the systemic velocity, but rarely redshifted and certainly not anywhere near this much.
Contrived models of \ion{He}{2} damped Ly$\alpha$ absorbers can be constructed at high redshift to account for an apparent line-center shift due to a strong red absorbing wing \citep{zheng08}, but no such model is possible with the low helium optical depth in the HS1700 sightline.
It is even more puzzling that the {\it FUSE} spectrum, discussed below, shows no evidence for this excess, despite otherwise agreeing well with the COS spectrum.

Therefore, we see no clear evidence for \ion{He}{2}~Ly$\alpha$ emission in HS1700.
This leaves no detections of this line at good S/N, and only a few at low S/N and very low resolution \citep{syphers09a}.
The lack of such emission despite predictions is puzzling, and may suggest that the models are overly simplistic.
Because \ion{He}{2} $\lambda304$ scatters resonantly within the broadline cloud, some of the emission will be absorbed in the \ion{H}{1} continuum, leaving only the surface layers to contribute to the observed line emission.
Also, the Cloudy photoionization modeling code does not properly handle Bowen resonance-fluorescence, where a \ion{He}{2}~Ly$\alpha$ photon excites \ion{O}{3} via a coincidence of transitions around 303.8~\AA\ \citep[][pp.~99--101]{bowen35,osterbrock06}.
However, Cloudy {\it does} destroy these photons based on the presence of \ion{O}{3}, which should suffice for our predictions. 
It is possible that the emission seen in HS1700 could come from metals.
Many models predict the blend of \ion{O}{3} lines around $\lambda$306~\AA\ to be the strongest metal emission anywhere near the break, although it is always predicted to be much weaker than \ion{He}{2}~Ly$\alpha$.
The number of quasars with good-quality spectra of the region near quasar-frame \ion{He}{2}~Ly$\alpha$ continues to increase, and will hopefully offer more clues about both helium and metal lines in the EUV.

\subsection{Comparison to FUSE}
\label{sec:fuse_comp}
HS1700 was observed by {\it FUSE} several times, most extensively in 2003 \citep{reimers06a}.
\citet{fechner06} performed the most detailed study of the {\it FUSE} data relating to the \ion{He}{2}~Ly$\alpha$ forest.
We present in Figure~\ref{fig:cos_fuse} an improved reduction of the {\it FUSE} spectrum (not previously published).
The data were reduced with CALFUSE 3.0.7 using night-only data (288~ks) and PHA channels 4--16, with customized background subtraction.
Individual exposures and segments were separately zero-corrected before coaddition.
The region 1081--1087~\AA\ is known to be unusually noisy, but we include it in our figure for comparison with COS.
\citet{fechner06} excised this region from their analysis, but they used an earlier {\it FUSE} reduction with worse background subtraction.

Despite using night-only data, we still find noticeable geocoronal contamination in the {\it FUSE} spectrum, in part because of the large 30$'' \times 30''$ (LWRS) aperture used.
This is nearly 200 times the area (and hence sky background) of the COS aperture, although both admit about the same amount of source flux.

{\it FUSE} had $R \sim $20,000, much higher than COS G140L ($R \sim 2500$), although it is possible that the resolution was degraded by pointing issues since the large aperture was used.
As this is ideally high enough to resolve the Ly$\alpha$ forest, efforts were made to do so with the only other \ion{He}{2} quasar observed with {\it FUSE}, HE2347 \citep{kriss01,zheng04b,shull04}.
Although the resolution may be good enough to do this reliably, the S/N likely is not, for either HE2347 or HS1700 \citep{fechner06}.
Unfortunately our COS data cannot resolve the forest, although a newly available bluer mode of the COS medium-resolution grating will allow such observations of some \ion{He}{2} quasars down to $z \simeq 2.5$.

\subsection{Comparison to STIS and the Intrinsic Spectral Index}
\label{sec:stis_comp}

\begin{figure}
\epsscale{1.2}
\plotone{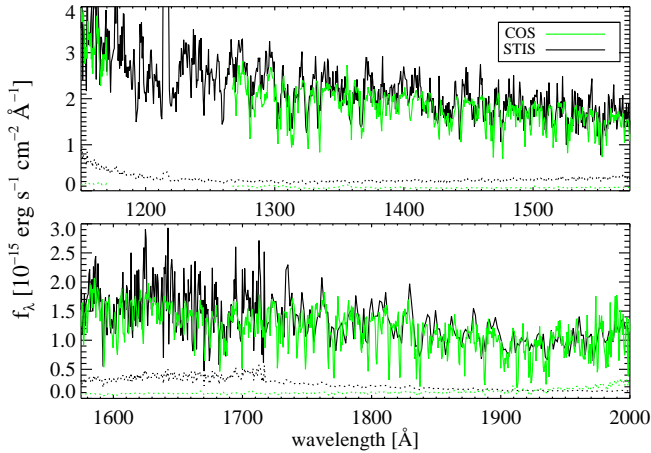}
\caption{The 2011 COS (green) and 2003 STIS (black) spectra of HS1700, overplotted, with error vectors plotted as dotted lines. The COS/G104L data are binned to one resolution element ($0.56$~\AA), while the STIS data are plotted at half a resolution element ($0.58$~\AA\ G140L, $1.54$~\AA\ G230L). Both are corrected for Galactic reddening of $E(B-V)=0.026$ using the extinction curve of \citet{fitzpatrick99}. The close agreement in flux between the two observations is somewhat surprising, given the historically highly variable nature of this source.}
\label{fig:cos_stis}
\end{figure}

HS1700 was observed with STIS in 2003, for 2245~s using G140L ($R \sim 1000$) and 2166~s using G230L ($R \sim 500$).
We plot this STIS data with our COS data superimposed in Figure~\ref{fig:cos_stis}.
HS1700 is known to be highly variable in the observed-frame FUV \citep{reimers05a}, but we see only small differences in flux and spectral index between the two observations.
Fitting a single power law below any strong pLLS Lyman breaks ($\lambda < 1900$~\AA), we find that the STIS data give $\alpha_{\nu}=0.086^{+0.07}_{-0.05}$ (for $f_{\nu} \propto \nu^{\alpha_{\nu}}$), while the COS data give $\alpha_{\nu}=-0.13^{+0.11}_{-0.14}$.

However, because HS1700 is one of the few quasars with continuous spectral observations from the \ion{He}{2} Ly$\alpha$ break to the \ion{H}{1} Lyman limit, it is possible to estimate all hydrogen absorption systems along the sightline.
We performed a spectrum fit to the STIS data including pLLS, using the redshifts and column densities of \citet{fechner06a} as initial parameters \citep[many of the column densities are ultimately from][]{vogel95}.
While the redshifts are derived from hydrogen and metal lines, and are therefore secure, the hydrogen column densities $N_{\rm HI}$ are less certain.
The low S/N and very low resolution ($R \sim 500$) of the STIS NUV data makes fits problematic.
Resolution is crucial even for measurements of the Lyman break, because the continua need to be well defined in a region with large amounts of contaminating absorption.

\begin{figure}
\epsscale{1.2}
\plotone{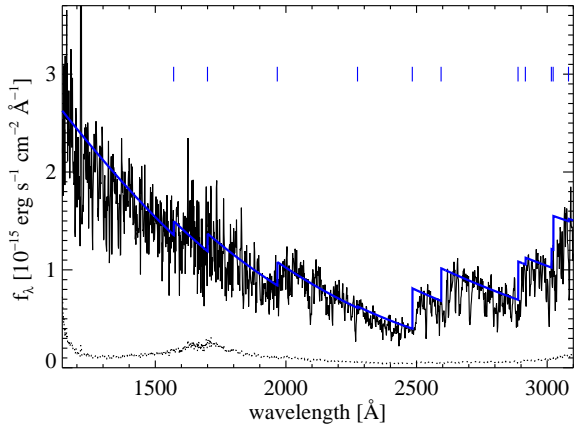}
\caption{The full STIS spectrum of HS1700 with an overplotted fit, including \ion{H}{1} pLLS. For clarity, only the Lyman breaks are shown (tick marks for those with $\log{N_{\rm HI}}>15.4$).  Higher Lyman-series lines were included for the fit, improving the alignment near Lyman breaks, where the apparent break is slightly redward of the Lyman limit because of line overlap. Data from G140L ($\lambda < 1715$~\AA) are binned to one resolution element ($\sim$$1.2$~\AA), and data from G230L are binned to half a resolution element ($\sim$$1.6$~\AA).}
\label{fig:stis_fit}
\end{figure}

The reasonably high S/N in the data allows us to detect pLLS  absorption edges  down to column densities $\log{N_{\rm HI}} \geq 15.2$, corresponding to optical depths $\tau_{\rm HI} = (0.01)(N_{\rm HI} / 10^{15.2}~{\rm cm}^{-2})$.
Our fit to the STIS data uses a single underlying power-law with 17 pLLS with $\log{N_{\rm HI}} \geq 15.2$, nine of which have $\log{N_{\rm HI}} \geq 16.2$ (see Figure~\ref{fig:stis_fit}).
For the most part we find column densities fairly close to those of \citet{fechner06a}, with the following exceptions.
For the $z=1.8450$ system, we use $\log{N_{\rm HI}}=16.80$. The \citet{fechner06a} value of $16.21$ was derived from fitting Lyman series lines, but is inconsistent with the strength of the break.
For the $z=2.1989$ system, we adopt $\log{N_{\rm HI}}=16.00$; while reported at $\log{N_{\rm HI}}=15.44$ in \citet{fechner06a}, a nontrivial break is helpful in the fit, although its column density is not well constrained.
At $z=2.2895$, we see no Lyman break seen associated with this system, despite the \citet{fechner06a} value of $\log{N_{\rm HI}}=16.00$ ($\tau_{\rm HI}=0.063$).

With the redshifts and column densities of the pLLS fixed, we then fit the COS data from the \ion{He}{2} break to the G140L instrumental cutoff at 2000~\AA.
We find the underlying EUV spectral indices to be $\alpha_{\nu}=-1.70$ (COS) and $\alpha_{\nu} = -1.62$ (STIS).
These values agree with each other, considering systematic uncertainties.
Quasars have widely varying spectral indices, but we note that this value is consistent with the average seen in radio-quiet {\it HST} EUV quasar spectra 
($\alpha_{\nu} = -1.57 \pm 0.17$, \citealt{telfer02}; $\alpha_{\nu}=-1.41\pm0.21$, \citealt{shull12a}), though not with the {\it FUSE} average 
\citep[$\alpha_{\nu} =- 0.56^{+0.38}_{-0.28}$,][]{scott04}.
The dramatic difference between the underlying spectral index and the apparent $\alpha_{\nu} \sim 0$ spectral index from a power-law fit to the COS data indicates the difficulty of discerning the true value without continuous spectral coverage from the Lyman limit.
HS1700 is, however, an extreme case with the number of pLLS it shows.  

\citet{giroux94} analyzed the metal-ion absorbers in these pLLS, showing that their abundances were inconsistent with photoionization equilibrium.
Instead, they proposed a multiphase medium with hot, collisionally ionized gas producing much of the high ionization states of the C, N, and O ions.
However, the \ion{H}{1}/\ion{He}{1} ratios crucial to the \citet{giroux94} analysis depend on FOS observations of \ion{He}{1} that are not entirely reliable, as we note above.
\citet{fechner06a}, using better data, fit the systems as photoionized only, but do not discuss the possibility of collisional ionization.

\begin{figure}
\epsscale{1.2}
\plotone{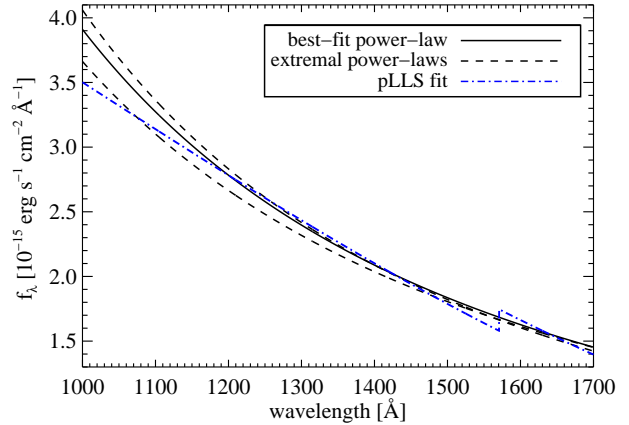}
\caption{A comparison of the pure power-law continuum fits to the best fit when including all the pLLS (though all but one of the Lyman breaks occurs at longer wavelengths than those shown here). The best-fit power-law is the solid black curve. The highest and lowest power-law fits for the 1000--1140~\AA\ region are dashed black lines (these are not the extremal fits at longer wavelengths). The best fit including all 17 of the $\log{N_{\rm HI}} > 15.2$ absorbers is the dot-dash blue curve. As one might expect, the fits are similar in the region where we have data, and differ noticeably only below the \ion{He}{2}~Ly$\alpha$ break ($1140$~\AA). However, even at 1000~\AA, the difference is not significant for calculating Gunn--Peterson optical depths; see text for details.}
\label{fig:extrapolated_continua}
\end{figure}

We do not necessarily regard the fit of HS1700 with pLLS as superior to the simple power-law fit for calculating Gunn--Peterson optical depths.
The two continuum extrapolations are compared in Figure~\ref{fig:extrapolated_continua}.
The pLLS fit does account for the non-power-law nature of the continuum, and should therefore indeed be superior, but there is nontrivial uncertainty in the column densities of the pLLS absorbers, and difficulty discerning the true continuum in data with low resolution and modest S/N \citep[where the true continuum is easily confused with emission lines;][]{shull12a}.
The high-transmission region near 1040~\AA\ matches the power-law continuum quite well (Fig.~\ref{fig:cos_fuse}), and the pLLS fit is slightly unphysical there, lying below the observed flux.
The noise is fairly high in this region, so this is not a definitive test, but it does show that the pLLS fit is not necessarily better.
The moderate-strength pLLS at $z<1$ can be included or not---the COS data do not strongly show them or rule them out.
Wiggles in the apparent continuum can be due to emission lines rather than rises and falls associated with pLLS.
In this case, because we have metal absorption data for these complexes \citep{fechner06a}, it makes it more likely that they are truly pLLS.
We do not see strong evidence for a spectral index break at 2000~\AA\ (observed), as suggested in \citet{fechner06}.

\section{THE IGM IONIZING BACKGROUND}
\label{sec:eta}

One of the more interesting pieces of information in a spectrum of \ion{He}{2}~Ly$\alpha$, when used in conjunction with an \ion{H}{1}~Ly$\alpha$ spectrum covering the same redshift, is the hardness of the EUV background.
The ratio is characterized by $\eta \equiv N_{\rm He II}/N_{\rm H I} \simeq 4 \times \tau_{\rm HeII}/\tau_{\rm HI}$.
While metal-line ratios from different ions can yield more detailed information, using helium gives the great advantage of almost complete coverage at every observable redshift.
Very high values of $\eta$ ($\gtrsim$500) are a sign of a soft ionizing field, dominated by stars, while low values suggest that the harder spectra of quasars dominates.

\begin{figure*}
\epsscale{0.8}
\plotone{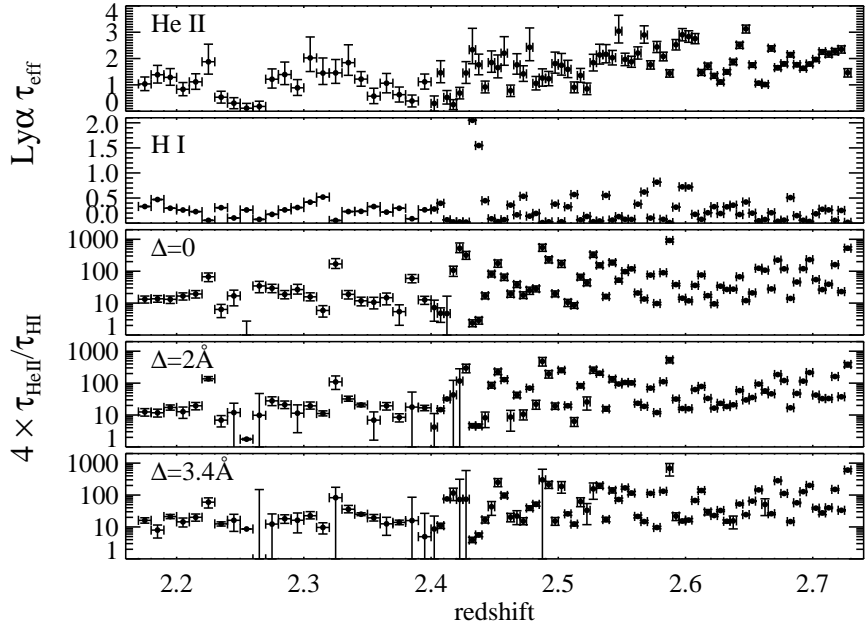}
\caption{Optical depths of \ion{He}{2} and \ion{H}{1} Ly$\alpha$ as a function of redshift, with $4 \times \tau_{\rm HeII}/ \tau_{\rm HI} \simeq \eta$ plotted in the lower three planels. The first $\eta$ panel uses the wavelengths as reported by CALCOS, while the second shifts by $\Delta = 2$~\AA\ and the third by $\Delta = 3.4$~\AA, reflecting uncertainty in the wavelength calibration. Our best estimate of the true shift is 2.0~\AA; see text for details. The data are binned to $\Delta z = 0.005$ (about 2.7 COS resolution elements) for $z \geq 2.4$, and $\Delta z = 0.01$ for $z<2.4$ due to lower S/N in this region. Errors shown are 68\% including Poisson uncertainty in FUV counts and continuum uncertainty. Uncertainty in $\tau_{\rm HI,eff}$ is smaller than the plotting symbols. The optical depth adjustments from accounting for contaminating (non-helium) absorption in the FUV spectrum are included, but the original values all lie within the plotted confidence intervals.}
\label{fig:eta}
\end{figure*}

Our plot of $\eta(z)$ is shown in Fig.~\ref{fig:eta}, but unfortunately low resolution and the wavelength calibration uncertainty do not allow us to draw strong conclusions about specific features.
The \ion{H}{1} data are convolved with the COS LSF prior to calculating $\tau_{\rm eff}$, although this has little impact because each redshift bin contains at least 2.7 COS resolution elements.

We find a median $\langle \eta \rangle_{\rm med}=31.7^{+1.5}_{-8.4}$ (68\% confidence interval from bootstrapping), with the 25$^{\rm th}$--75$^{\rm th}$ percentile range being $\eta = 12$--128.  For the mean we find $\langle \eta \rangle_{\rm mean} = 73^{+10}_{-11}$.
\citet{fechner06} found $\langle \eta \rangle_{\rm med} = 51$ in the HS1700 {\it FUSE} data, and $\langle \eta \rangle_{\rm med} = 79$ when they restrict to a subset of the data where they argue $\tau_{\rm HI}$ is more robustly measured.
Along the HE2347 sightline, $\eta$ at $z=2.4$--2.73 was found to be $\langle \eta \rangle_{\rm med} = 33$ \citep{shull10}.  The mean was $\langle \eta \rangle_{\rm mean} = 78 \pm 7$ in \citet{kriss01} and similar in other works \citep[][although the latter quotes extremely large uncertainties]{zheng04b,shull10}.

Most of the $z=2.2$--2.7 IGM shows $\eta$ values consistent with quasars being the dominant source of ionizing photons, although some regions show softer backgrounds ($\eta \gtrsim 500$) that require strong filtering or local dominance by softer sources such as star-forming galaxies.  Such a hard background on average is expected in the \ion{He}{2} post-reionization epoch, where there is no longer strong absorption at the 4~ryd helium edge, as there was at higher redshift.
\citet{muzahid11} claim that the average IGM region has a fairly high $\eta$ based on the HE2347 sightline, and contend that low values ($\eta \lesssim 40$) may be signs of gas that is not photoionized.  The contribution of collisionally ionized \ion{He}{2} was seen in simulations \citep{shull10}.  
Nonetheless, quasars are fully capable of producing lower $\eta$ values, and indeed one might expect $\eta \sim 30$ for the {\it average} quasar spectrum filtered through an optically thin medium \citep{fardal98}, aside from the many quasars that have spectra harder than average.
In addition, median values of $\eta$ found here and some other works \citep[e.g.,][]{shull10} do not support the claim of high $\eta$ being common after helium reionization, although we caution that we are using somewhat lower resolution data than \citet{muzahid11}.
With $\tau_{\rm eff}$ this makes a difference, as higher-transmission regions dominate, but as this affects both hydrogen and helium spectra, it is unclear which way this would skew $\eta$.  Data in the \ion{He}{2}~Ly$\alpha$ forest at good resolution {\it and} good S/N is the only way to conclusively examine this, but as yet no such data exist.  By considering the ``sawtooth-modulated spectrum"  produced by higher Lyman lines of  \ion{He}{2},  \citet{madau09} predict a \ion{He}{2}/\ion{H}{1} ratio of 
$\eta \approx 35$ for optically thin filtering. 

Because $\eta$ fluctuates strongly on fairly short scales \citep{shull10}, its measurement is sensitive to systematic velocity offsets between the UV and the optical data.
While we can confirm these averages for the post-\ion{He}{2} reionization epoch, unfortunately, due to a larger-than-expected uncertainty in the wavelength calibration of the COS data, we are unable to speak about details in the $\eta$ fluctuations along the HS1700 sightline.

Last we note a small but nonzero chance that there is a systematically incorrect estimation of our optical depths due to the possibility of wavelength calibration uncertainty in the sensitivity curve.
In our determination of the optical depths, we use the sensitivity curve of segment B data to predict the number of counts expected for an unabsorbed spectrum.
For $\lambda \gtrsim 1080$~\AA, the effective area of the COS detector changes rapidly \citep{mccandliss10}, and a small shift in wavelength could have an impact.
For example, a wavelength shift of 2~\AA\ leads to an optical depth shift of $\Delta \tau \simeq 0.1$.
This shift, while not entirely trivial, is small compared to the values and other sources of uncertainty, so even in the unlikely case that such a large wavelength shift were present in the calibration data, its impact on our science would be minimal.

\section{PROXIMITY EFFECTS}  
\label{sec:proximity}

\subsection{Line-of-Sight Proximity Effect}
\label{sec:LOS_prox}

\begin{figure*}
\epsscale{0.8}
\plotone{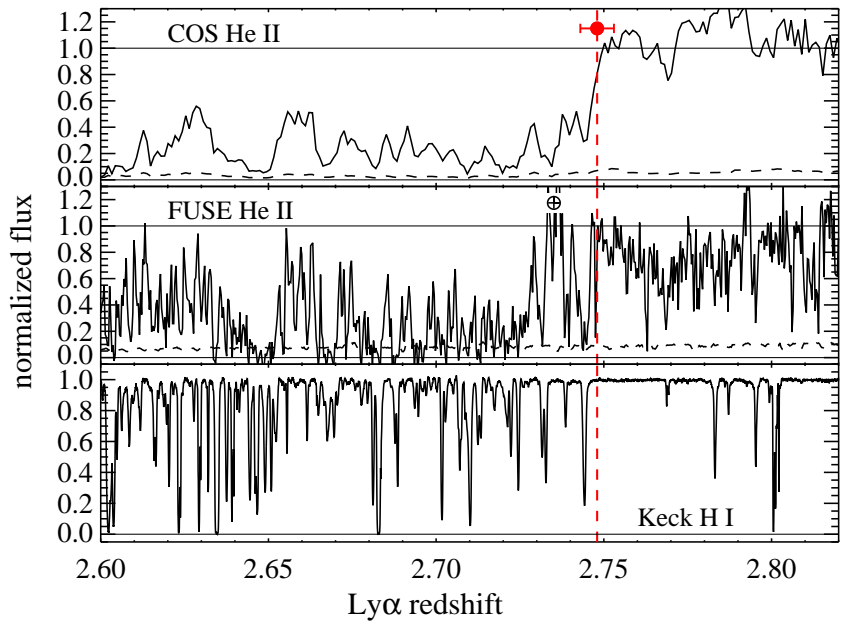}
\caption{Normalized flux for the COS, {\it FUSE}, and Keck spectra, shifted to redshift space for \ion{He}{2} and \ion{H}{1} Ly$\alpha$. The redshift of HS1700 is marked with the dashed vertical line, with approximate redshift uncertainty shown in the COS panel. Error vectors are overplotted as dashed lines for the UV data. COS data are binned to $\simeq$85~km~s$^{-1}$ (just over half a resolution element), FUSE data are binned to $\simeq$27~km~s$^{-1}$ (1--2 resolution elements), and Keck data are binned to $\simeq$7~km~s$^{-1}$ (one resolution element). The {\it FUSE} spectrum is contaminated by geocoronal \ion{N}{1}~$\lambda$1134, which is marked.}
\label{fig:zoomin}
\end{figure*}

Uncertainty in the systemic redshift of HS1700 has been a major source of uncertainty in interpreting the line-of-sight proximity effect.
While nontrivial uncertainty remains, the redshift is now determined well enough to address this issue.
Unfortunately, the COS data are of low resolution, and the break appears to be abrupt.
We therefore consider the COS data in conjunction with the {\it FUSE} and Keck \ion{H}{1} data, in Figure~\ref{fig:zoomin}.
Wavelength calibration uncertainty (section \ref{sec:wavcal}) is larger than systemic redshift uncertainty for evaluating the proximity effect, but the suggested zone is large enough---and well enough aligned with {\it FUSE} and Keck features---that this does not affect our results.

At first glance, the {\it FUSE} data suggest there is essentially no proximity effect at all.
This would be unexpected for such a luminous quasar, although not unprecedented as no proximity zone is seen in HE2347 either \citep[but this is complicated by strong intrinsic absorption;][]{fechner04,shull10}.
However, the apparent break in the {\it FUSE} spectrum is associated with a fairly strong absorber seen in \ion{H}{1} as well.
It is quite plausible that there is a \ion{He}{3} zone surrounding this quasar out as far as the very high transmission in hydrogen is seen ($\sim$14--24 comoving Mpc), interrupted by surviving dense absorbers.
The COS spectrum, which does not suffer from geocoronal \ion{N}{1}~$\lambda$1134 contamination like the {\it FUSE} spectrum does, supports this possibility.
If the transmission peak near $z \simeq 2.73$ is indeed associated with the proximity zone, then the highest-redshift point in Fig.~\ref{fig:tau_evolution} should be shifted up slightly, because some of this transmission lies in that bin.

\subsection{Transverse Proximity Quasars}
\label{sec:transverse_prox}

The transmission observed in the \ion{He}{2}~Ly$\alpha$ forest is in part a reflection of the overall state of helium in the IGM at a given redshift, a measurement of the global ionizing background that evolves during helium reionization.
However, it also can depend on local ionizing sources.
In particular, a quasar that is near the sightline can affect the ionization state of the gas along the sightline, leading to a transverse proximity effect.
Measurements of this effect can give information about quasars, notably lifetimes and beaming angles \citep{furlanetto11,lu11}.

Measurement of increased transmission due to the transverse proximity effect on the \ion{H}{1}~Ly$\alpha$ forest has been unsuccessful \citep[e.g.,][]{kirkman08}, likely due to factors such as quasars existing in regions that are denser than average, counteracting their increased ionizing flux.
It has proved much more fruitful to look at ion ratios, which give information on spectral hardness and break the degeneracy between ionizing flux and IGM density.
Metal systems can be used for such ratios \citep{goncalves08}, but the vast majority of the path length through the Ly$\alpha$ forest has no observable metal systems.
It has therefore been rewarding to compare the hydrogen and helium Ly$\alpha$ forests, which gives a measurement of spectral hardness that can be used anywhere.
Several detections of transverse proximity quasars have been made this way, for the sightlines to Q0302$-$003 \citep{jakobsen03,worseck06} and HE2347 \citep{worseck07}.

\begin{figure*}
\epsscale{0.9}
\plotone{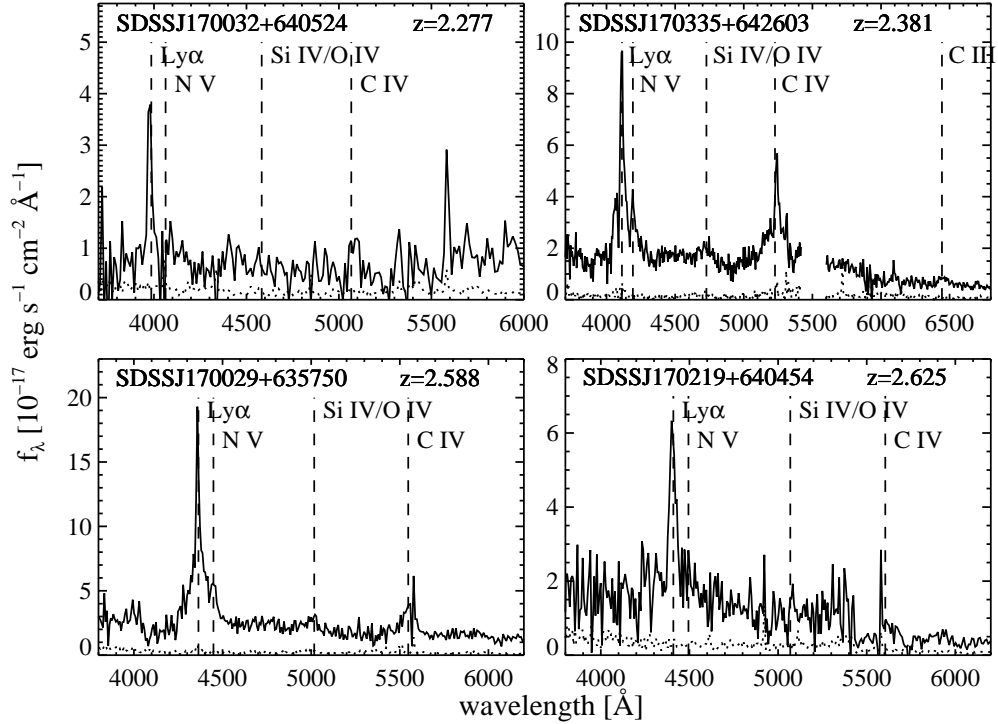}
\caption{Spectra from APO/DIS verifying the nature and redshifts of potential transverse proximity quasars (high redshift QSOs near the HS1700 sightline with $z_{\rm QSO} < z_{\rm HS1700}$). Spectra from red and blue detectors are joined at about 5400~\AA; this is particularly obvious for the highest-redshift object, where the detection of the \ion{C}{4} line must be judged using red-side data only. Further details given in Table~\ref{tab:trans_obs}.}
\label{fig:trans_prox}
\end{figure*}

We present the first survey for high-redshift quasars near the HS1700 sightline (Figure~\ref{fig:trans_prox}, Table~\ref{tab:trans_obs}).
The field near HS1700 is covered by SDSS imaging, with good-quality $u$$g$$r$$i$$z$ data.
Unfortunately, the region near $z \sim 2.7$ is a known redshift desert for SDSS \citep{richards09}, due to quasars overlapping the stellar locus in color-color space.
Efforts have been made to construct photometric quasar catalogs nonetheless, and in our selection we draw on two of these, \citet{richards09} and \citet{bovy11}.
Because the targets observed spectroscopically were chosen for a variety of factors, we emphasize that this survey has considered most ``good'' candidates within $\sim$20$'$ ($\sim$10~proper Mpc) of the sightline with $r < 22$, but it is not complete.

We note that more efficient differentiation of quasars and stars can be achieved with many epochs of imaging allowing variability selection \citep[e.g.,][]{macleod11}, or with IR bands for those quasars bright enough to be detected in {\it WISE} \citep{wu12}.
Slitless spectroscopy is difficult for large fields in the optical, due to crowding, but it has been successfully used to identify transverse proximity quasars in some southern fields \citep{worseck08}.

Our spectra of candidate transverse proximity quasars are shown in Figure~\ref{fig:trans_prox}.
To verify candidate quasars and obtain redshifts, we used the Dual Imaging Spectrograph (DIS) on the 3.5m telescope at Apache Point Observatory (APO).
DIS has a blue side and a red side, with a constant dispersion of $1.8$~\AA~pixel$^{-1}$ in the blue and $2.3$~\AA~pixel$^{-1}$ in the red, with two pixels per resolution element.
We use the blue side for $\lambda < 5425$~\AA, and the red side longward of this.
Independently calibrated, the two sides normally match fluxes seamlessly, but factors such as a poor trace of a faint spectrum can cause matching problems such as the one seen in Figure~\ref{fig:trans_prox} for SDSSJ170219+640454.

\begin{deluxetable*}{llllcccl}
\tablecolumns{8}
\tablewidth{0pc}
\tabletypesize{\footnotesize}
\tablecaption{APO/DIS Observations of Potential Transverse Proximity Quasars}
\tablehead{
\colhead{Name} & \colhead{R.A.} & \colhead{Decl.} & \colhead{Redshift} & \colhead{Impact Parameter} & \colhead{$r$\tablenotemark{a}} & \colhead{Exp. Time} & \colhead{Obs. Date} \\
\colhead{} & \colhead{(J2000)} & \colhead{(J2000)} & \colhead{} & \colhead{[proper Mpc]} & \colhead{[mag]} & \colhead{[min]} & \colhead{(UT)}}
\startdata
SDSSJ170032.86+640524.8 & 255.13694 & 64.09025 & $2.277 \pm 0.012$\tablenotemark{b} & 3.72 & 21.7 & 60 & 2011.08.27 \\
SDSSJ170335.96+642603.8 & 255.89987 & 64.43439 & $2.381 \pm 0.003$ & 10.9\phn\phn & 21.7 & 90 & 2011.05.31 \\ 
SDSSJ170029.48+635750.9 & 255.12286 & 63.96415 & $2.588 \pm 0.014$ & 7.40 & 21.2 & 60 & 2012.05.29 \\ 
SDSSJ170219.00+640454.5 & 255.57921 & 64.08182 & $2.625 \pm 0.005 $ & 5.64 & 20.9 & 100 & 2012.05.16 \\ 
\enddata
\tablenotetext{a}{~From SDSS DR8 \citep{eisenstein11}. These are not necessarily the magnitudes used in constructing the photometric catalogs---for example, in DR7 \citep[used by][]{richards09} the $z=2.381$ quasar was $0.8$ magnitudes brighter in $r$, due to some combination of quasar variability and altered data processing.}
\tablenotetext{b}{~Line identifications are tentative; see text for details.}
\label{tab:trans_obs}
\end{deluxetable*}

The identification of SDSSJ170032+640524 is tentative.
There is only one very clear emission line in this spectrum, and the redshift depends on identifying this as Ly$\alpha$.
\ion{C}{4} 1549 does appear to be present at a consistent redshift, but no other lines are seen.
If the strong emission line is instead \ion{Mg}{2}, we should see H$\beta$ in the spectrum, but we do not.
Spikes near 5600~\AA\ and 6300~\AA\ are narrow---consistent with purely instrumental broadening ($\sim$250~km~s$^{-1}$), and thus likely noise---while the identified emission line has a FWHM of $\sim$1300~km~s$^{-1}$.
This would make it narrow for a quasar, but much too broad to be from a star-forming galaxy.

SDSSJ170335+642603, at $z=2.381$, is unfortunately of minimal observational interest for transverse proximity effects.
At that redshift, \ion{He}{2}~Ly$\alpha$ is observed at 1027~\AA---right on top of geocoronal \ion{H}{1}~Ly$\beta$, a strong contaminating feature in our COS spectrum.

There is a significant protocluster of galaxies at $z \simeq 2.30$ within a few arcminutes of the sightline; this protocluster should generate a soft UV background, with many star-forming galaxies but no reported AGN \citep{steidel05}.
This could therefore potentially affect \ion{H}{1}~Ly$\alpha$ transmission in the sightline, but not \ion{He}{2}~Ly$\alpha$ transmission.
However, we see no evidence for a soft photoionization effect (Fig.~\ref{fig:eta}).

\section{CONCLUSIONS}
\label{sec:conclusion}

We have presented a new COS spectrum of the bright \ion{He}{2} quasar HS1700.
This spectrum has the highest S/N and resolution of any UV spectrum of this quasar in the region $1170 \lesssim \lambda \lesssim 2000$~\AA, and higher S/N and much better background subtraction than the {\it FUSE} spectrum at shorter wavelengths.
This allows us to accurately calculate $\tau_{\rm eff}(z)$ for this sightline, finding values significantly higher than previously reported, at in agreement with values found for the HE2347 sightline at the same redshifts.
The \ion{He}{2} post-reionization epoch is characterized by steadily falling effective optical depths, and a UV background that is varying but consistent with being dominated by the hard spectrum of quasars.

HS1700 was one of the few \ion{He}{2} quasars with a claimed detection of \ion{He}{2}~Ly$\alpha$ emission, and by far the brightest.
Our COS spectrum does indeed show an excess of flux near the \ion{He}{2}~Ly$\alpha$ break, significantly above any reasonable continuum fit.
However, the apparent peak of this flux is about 3000~km~s$^{-1}$ away from \ion{He}{2}~Ly$\alpha$ at the quasar rest-frame, even when using the much higher and more accurate systemic redshift of $z_{\rm HS1700}=2.75$.

Transverse proximity quasars are difficult to detect, but one of the most promising methods has been to look at variations of the spectral hardness using $\eta$.
We present four candidate proximity quasars along the HS1700 sightline.
Although none are clearly detected in the present data by $\eta$ variations, two are at high enough redshift that they could be observed at $R > 10$,000 with a newly available COS mode.
This resolution, combined with the much greater wavelength calibration certainty compared to the present G140L data, could definitively answer whether or not they are observed in the transverse proximity effect.
Either answer has interesting implications for quasar physics.

Background determination is one of the largest sources of uncertainty for studies of effective optical depth.
In this paper we present and use for the first time a new method of COS background subtraction that substantially improves on the pipeline strategy.
This method will be even more important when looking both at fainter targets, and crucially, prior to the end of helium reionization, when the optical depths are large.

\acknowledgments

We thank Rob Simcoe for providing the Keck spectrum of HS1700, Jerry Kriss and Jennifer Scott for providing their final reduction of the {\it FUSE} spectrum, and Justin Ely for useful conversations regarding the COS background.

This work was supported by NASA grants NNX08AC146 and NAS5-98043 to the University of Colorado at Boulder.

{\it Facilities:} \facility{HST (COS)}, \facility{APO (DIS)}

\appendix

\section{Determining the COS FUV Background}
\label{app:bg}

The background of COS is dominated by dark current from the detector, except at specific wavelengths where geocoronal line emission dominates.
Scattered and Zodiacal light is negligible \citep{syphers12}.
The current CALCOS (2.18.5) method for background determination uses two windows offset from the primary science aperture (PSA) in the cross-dispersion ($y$) direction, with a running average over 100 pixels\footnote{We refer throughout this discussion to pixels, which unless otherwise specified means the two-dimensional pixels created by digitizing the output of the cross delay line detector. While the COS detector itself does not have physical pixels, they are fundamental in the data the observer receives.} in the dispersion ($x$) direction.
The drawbacks of this method are discussed extensively in \citet{syphers12}, and here we present the first implementation of the improved method suggested in that work.

The primary problem is so-called $y$-dip, where counts in the PSA have lower pulse-height amplitudes (PHAs) than those in background regions (see Fig.~6 in \citealt{syphers12}).
Because of detector thresholds and PHA cuts, this means the background in the PSA is actually noticeably lower than the background inferred from regions offset in $y$.
(We note that this problem is lessened but by no means removed if no PHA filtering is imposed, and such inclusive analysis adds a large amount of noise.)
On July 23, 2012, the PSA was shifted to a new lifetime position on the detector, but $y$-dip will again become an issue as the new position is used more.
Also, an {\it underestimate} of the background could occur if background windows overlap regions previously affected by $y$-dip.

\citet{syphers12} suggest using data from the COS FUV dark monitoring programs (GO 11895, 12423, and 12716) to precisely characterize the dark current in the PSA, and use this for background subtraction.
Although ideally the dark would be recorded without binning, the very low dark current precludes this.
The rate at $2 \leq {\rm PHA} \leq 30$ is $\simeq 2 \times 10^{-6}$~counts~s$^{-1}$~pixel$^{-1}$, and the dark monitoring programs collect data fairly infrequently---once a week for about 27~ks per month for 11895 (September 2009 to October 2010) and once every two weeks for 13~ks per month for the more recent programs.
At this pace it would require years of collection to obtain on average a single background count per pixel.
The $y$-dip evolves noticeably with time, because it is caused by burn-in of the detector, so averaging data over too-long intervals is problematic.
Absent more frequent dark monitoring, we must both average moderately long time intervals and smooth the background.

We exclude dark exposures with unusually high overall count rates (we used the limit $6 \times 10^{-6}$~counts~s$^{-1}$~pixel$^{-1}$, but caution that the average background does vary noticeably with the solar cycle).
There are two categories of individual pixels with excessive counts: transient hot spots and consistent hot spots.
Individual exposures are checked for transient hot pixels, and we exclude any pixel with an excursion that would be expected $<1$\% of the time on the detector, assuming Poisson noise.
Consistent hot pixels also exist and are excluded.
These show up as very improbable excursions on the coadded master dark, but on any individual exposure they are often probable enough to not be excluded.
A true pixel-by-pixel master dark would include such elevated rates, but our smoothed version does not, as that would contaminate surrounding pixels.
In addition, when working in regions that are very sensitive to background rates, we recommend excluding regions known to have elevated backgrounds (data quality flag 32).

We also impose a PHA cut on our data that is more stringent than the pipeline standard.
The CALCOS pipeline by default includes 2~$\leq$~PHA~$\leq$~30 (recorded PHA ranges from 0 to 31).
However, the PHA distribution for sources is sharply peaked at lower PHA, while the PHA distribution for dark counts is much flatter.
For the segment A, tests with HS1700 showed that source counts contributed significantly only to the bins 1~$\leq$~PHA~$\leq$~14, and there were no detectable source counts for PHA~$\geq$~16 (whereas about 30\% of dark counts have PHA~$\geq$~16).
The detection of source counts in the PHA~$=$~15 bin is marginal, and in any case $< 0.07$\% of source counts have this PHA.
Counts with PHA~$=$~1 are a more subtle case, as $\simeq$70\% of them are source counts, but they are not included in the flux calibration.
We choose to neglect these counts---only $\simeq$$0.5$\% of source counts lie in this bin, and the important thing for our optical depth analysis is to predict source counts for an unabsorbed spectrum (using the flux calibration to convert from extrapolated flux to counts) and measure them in exactly the same way.
Segment B is similar, although at slightly higher PHA due to the higher voltage of that segment.
There are no detectable source counts with PHA~$=$~2--3, but nontrivial numbers have PHA~$=$~15--16.
(We note that there {\it are} substantial source counts with lower PHA in those portions of the detector strongly affected by gain sag due to geocoronal Ly$\alpha$, but all such regions lie below the Galactic cutoff of 912~\AA\ for G140L data.)
Our analysis of the HS1700 data is therefore restricted to 2~$\leq$~PHA~$\leq$~14 on segment A, and 4~$\leq$~PHA~$\leq$~16 on segment B.
We do recommend analyzing a specific data set prior to making PHA cuts, because the PHA distribution has several time dependencies, notably including loss of sensitivity in the exposed portion of the detector, shifts to a new lifetime position (as was done in July 2012), or changes in the high-voltage settings.

The spectrum extraction width (in the cross-dispersion direction) also contributes substantially to how much background is included.
The default extraction width for COS G140L data is 57 pixels in $y$, but the spectrum width is much smaller than this, and therefore many unnecessary dark counts are included.
For the HS1700 data analyzed here, we use an extraction width of 30 pixels for segment A, and 35 pixels for segment B.
This is fairly generous, as $>$95\% of all source counts are contained within 20 pixels over most of the detector, but we err on the side of being inclusive because the flux calibration will be off if too many source counts are cut out.
Also, the spectrum widens slightly at detector edges, including high wavelengths on the segment A ($\lambda \gtrsim 1900$~\AA) and the usable portion of the segment B ($1000 \lesssim \lambda \lesssim 1175$~\AA).
With our chosen widths, we preserve essentially all source counts for most of the spectrum, and $>$98\% at all wavelengths.

We directly verify flux accuracy by comparing standard extracted spectra to our extracted versions, and finding essentially identical flux levels in regions where the background is negligible.
With both PHA cuts and a narrower extraction width, we reduce our background to about 40\% of what it would be in a default extraction, without appreciably losing source counts.

\end{document}